\documentclass[12pt]{iopart}
\usepackage{iopams}
\usepackage{setstack}
\usepackage{subcaption,graphicx}
\usepackage{xcolor}

\newcommand{\colour}[1]{\textcolor{black}{#1}}

\begin{document}

\title[Elliptical cavities for intensity and position sensitive beam measurements]{Conceptual design of elliptical cavities for intensity and position sensitive beam measurements in storage rings}

\author{M S Sanjari$^{1,2}$, X Chen$^{1,3}$, P H\"ulsmann$^1$, Yu A Litvinov$^{1,3}$, F Nolden$^1$, J Piotrowski$^{1,3,4}$, M Steck$^1$, Th St\"ohlker$^{1,5}$}
\address{$^1$GSI Helmholtzzentrum f\"{u}r Schwerionenforschung, 64291 Darmstadt.}
\address{$^2$ExtreMe Matter Institute EMMI, 64291 Darmstadt.}
\address{$^3$Ruprecht-Karls-Universit\"{a}t Heidelberg, 69117 Heidelberg.}
\address{$^4$AGH University of Science and Technology, 30-059 Krakow.}
\address{$^5$Helmholtz Institut Jena}
\ead{s.sanjari@gsi.de}

\begin{abstract}

Position sensitive beam monitors are indispensable for the beam diagnostics in storage rings. Apart from their applications in the measurements of beam parameters, they can be used in non-destructive in-ring decay studies of radioactive ion beams as well as  enhancing precision in the isochronous mass measurement technique. In this work, we introduce a novel approach based on cavities with elliptical cross-section, in order to compensate \colour{the limitations of known designs for the application} in ion storage rings. The design is aimed primarily for future heavy ion storage rings of the FAIR project. The conceptual design is discussed together with simulation results.
\end{abstract}



\maketitle

\section{Introduction}
Cavity based BPM designs have advantages when high resolution and sensitivity are desired, albeit at the cost of narrow-band operation. First ideas date back as early as 1960s. In the literature, most successful designs were utilized in machines with small beam pipe aperture sizes. This leads to precision use of higher order modes, such as the dipole mode in the cavity, for the purpose of position determination \colour{(see e.g. \cite{kim2012})}. Some of these designs have aimed for similar constructions in ion storage rings (an overview can be found in \cite{chen2014-gsi}). But since the figures of merit of cavity resonant pick-ups degrade as a result of large beam pipe apertures and low beam intensities, a situation often found in heavy ion storage rings, a more suitable method is needed in order to achieve high sensitivities.

\section{Theory of operation}

\subsection{Schottky noise analysis}

The theory of Schottky noise analysis in storage rings is very well explained in the literature \cite{chattopadhyay1984}. Schottky noise can be used to extract several important beam parameters. Longitudinal signals provide measures of revolution frequency, momentum spread and intensity of the beam, while transversal signals can be used to extract information on tune and chromaticity. Longitudinal Schottky signals can also be used for in-ring mass and lifetime measurements of radioactive ion beams \cite{bosch2013}. \colour{The} power spectral density of the Schottky signals contains repeated bands around multiples of revolution harmonics that grow in width with increasing frequency. Since these bands essentially contain the same information on the beam, \colour{and} as long as these bands do not overlap, higher harmonics are preferable for higher frequency resolutions, provided the signal is mixed down to base band for analysis.

\subsection{Microwave Cavities as Schottky pick-ups}

The eigenmodes of microwave cavities can be modelled by using ideal RLC elements, where an electric current oscillates at eigenfrequency $f_\nu$ with quality factor $Q_\nu$ and shunt impedance $R_{sh,\nu}=U_\nu^2/P_{diss, \nu}$\colour{, where $U_\nu$ is the induced voltage across the gap in the cavity and $P_{diss, \nu}$ is the dissipated power in each eigenmode $\nu$ respectively}. Often the material independent \textit{characteristic impedance} is used in order to be able to compare different cavity structures:
\begin{equation}
\left(\frac{R_{sh}}{Q}\right)_\nu=\widehat{\left(\frac{R_{sh}}{Q}\right)}_\nu  \Lambda_\nu(\beta)^2
\label{eqn:roq}
\end{equation}
where $\Lambda_\nu(\beta)$ is the so called \emph{transit time factor} that is a function of the relativistic $\beta$ of the beam for each mode. The hat notation shows the ideal or \textit{frozen} characteristic impedance, namely for a cavity with zero length and for a beam travelling with the speed of light. 

A beam passing through a cavity excites oscillating fields. The resulting standing waves can be extracted out of the cavity by using proper couplers. At critical coupling the average output power of a single particle at the harmonic $m$ is \cite{sanjari2013}
\begin{equation}
\langle P_{out}\rangle|_{mf_r}=\langle P_{diss}\rangle|_{mf_r}=(Ze)^2f_r^2\ \widehat{R_{sh,\nu}}\ \Lambda(\beta)^2
\end{equation}

The use of RF cavities as pick-ups allows for sensitive detection of particles whenever one of the eigen-frequencies of the cavity matches with a harmonic of the beam. For intensity measurements, a longitudinally sensitive detector can be designed using a circularly cylindrical shallow pillbox, where an electromagnetic field oscillates at its fundamental oscillating eigenmode TM010. By designing proper geometrical dimensions and a tuning mechanism for the eigenfrequency, the above requirements for sensitive particle detection can easily be met.

\section{Transversal sensitivity}

\subsection{The R/Q map}
The longitudinal characteristic impedance R/Q is an integrated quantity as seen by a particle beam along the axis of the cavity resonator (z axis). Nevertheless, its value depends not only on the transit time factor as seen in equation (\ref{eqn:roq}). Due to the distribution (pattern) of the z component of the electric field in every specified mode, it also depends on the transversal position of the beam \cite{whittum1999}. Since the transit time factor can be determined numerically for a given beam species and energy, in the following we concentrate on the \textit{frozen} characteristic impedance, which is independent from the beam velocity.

A plot of the longitudinal characteristic impedance in 3D results in an \emph{R/Q map} which can be used to indicate the sensitivity of each modes for a given transversal offset. A circularly cylindrical shallow cavity shows a relatively flat R/Q map around its center which is also the location of the beam pipe, so this construction can be used for highly sensitive beam intensity measurements (see e.g. \cite{nolden2011} and \cite{kienle2013} for single particle sensitivity).

\subsection{Generalization to elliptical cross-section}
For the transverse sensitivity, a position dependent R/Q map is required  where a linear dependency is needed only along one transversal axis and ideally no variation along the other. Application of two such detectors placed sequentially in right angles on the beam axis will provide information on the whole transversal plane.
\begin{figure}[htb]
\centering
  \centering
  \subcaptionbox{\label{fig:circ_mode1e}M1: cir. TM$_{010}$}{\includegraphics[width=0.3\textwidth]{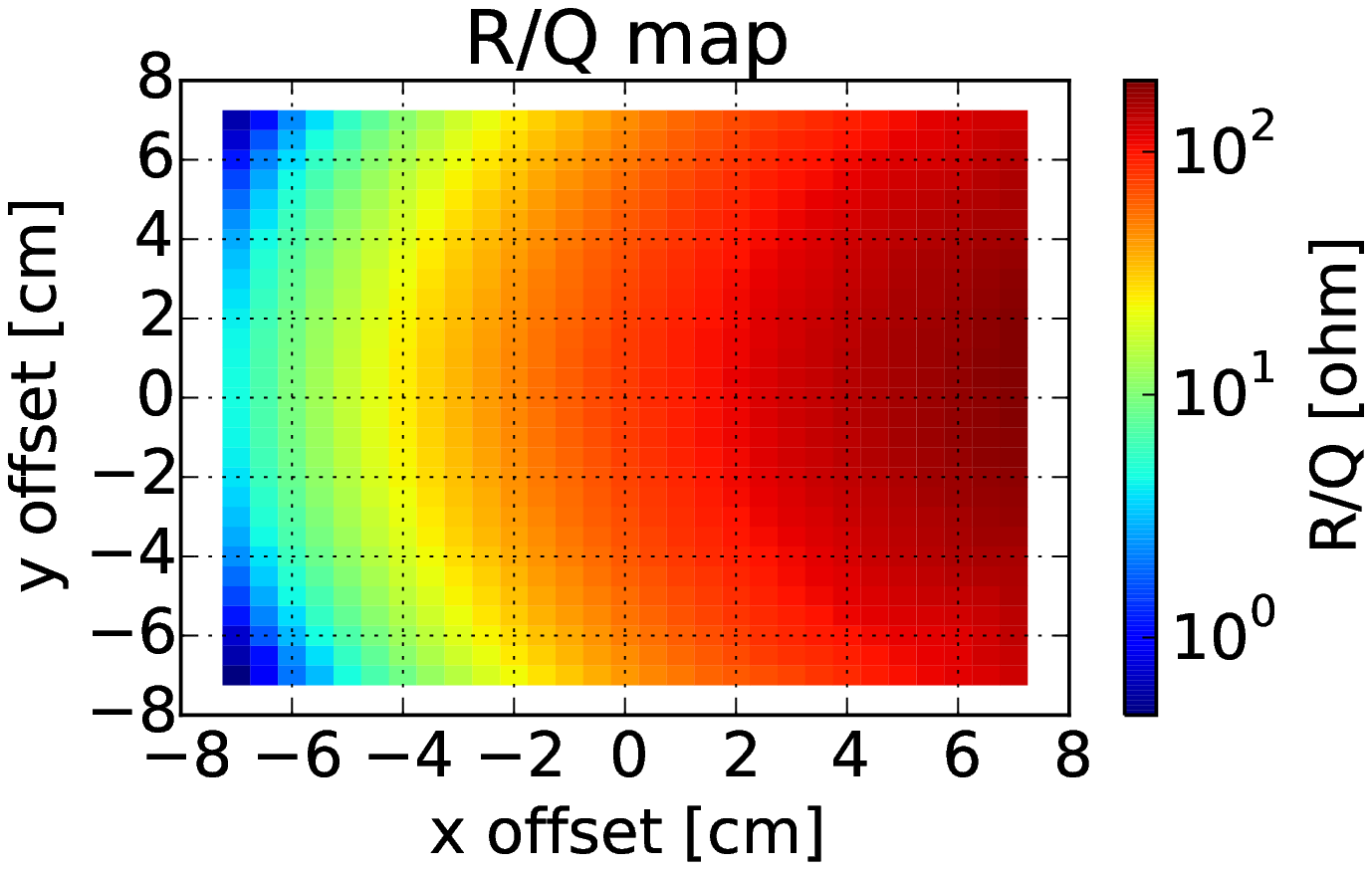}}  
  \subcaptionbox{\label{fig:circ_mode2e}M2: cir. TM$_{110}$}{\includegraphics[width=0.3\textwidth]{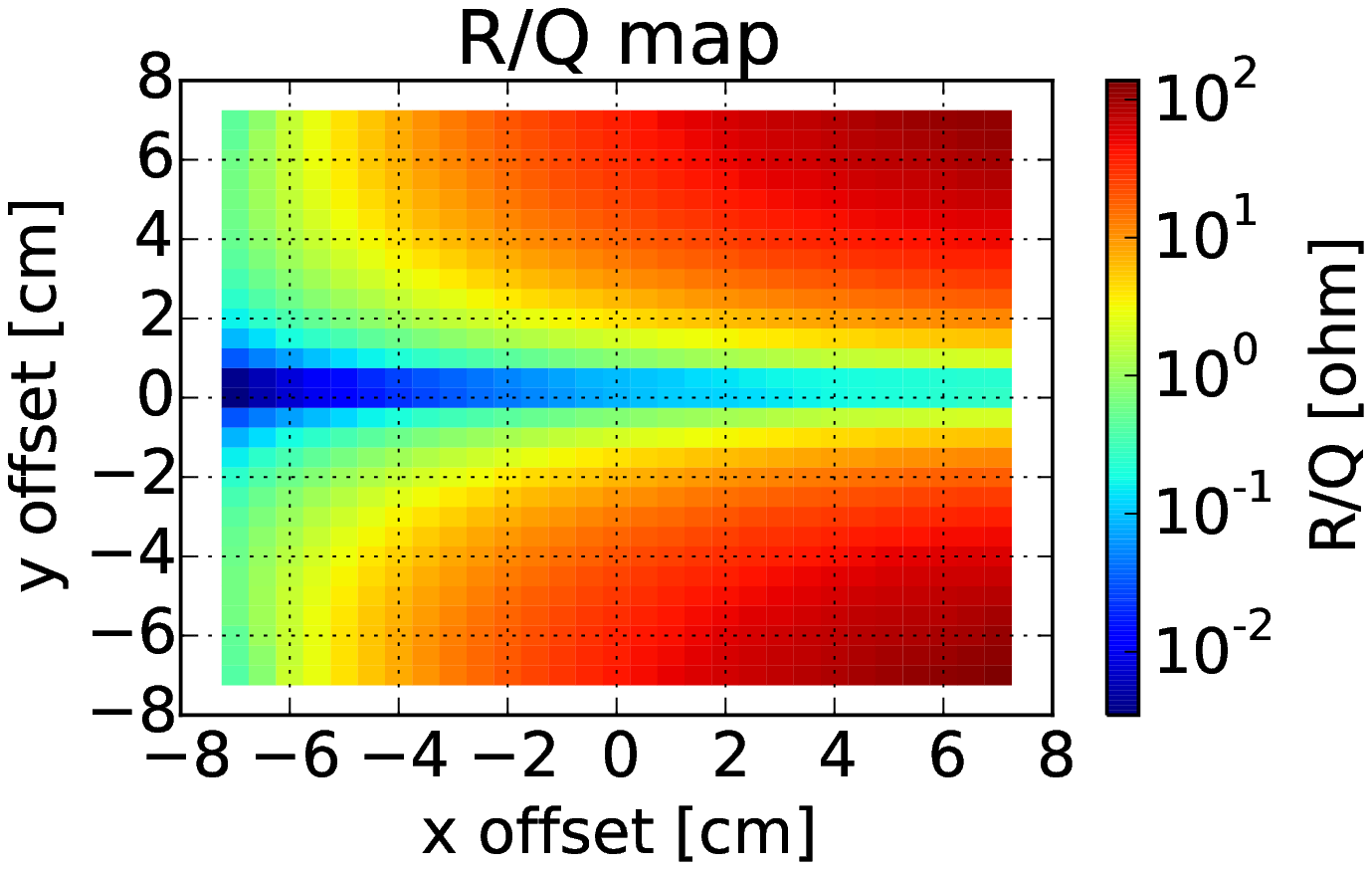}}  
    \subcaptionbox{\label{fig:circ_mode3e}M3: cir. TM$_{110}$ 2nd pol.}{\includegraphics[width=0.3\textwidth]{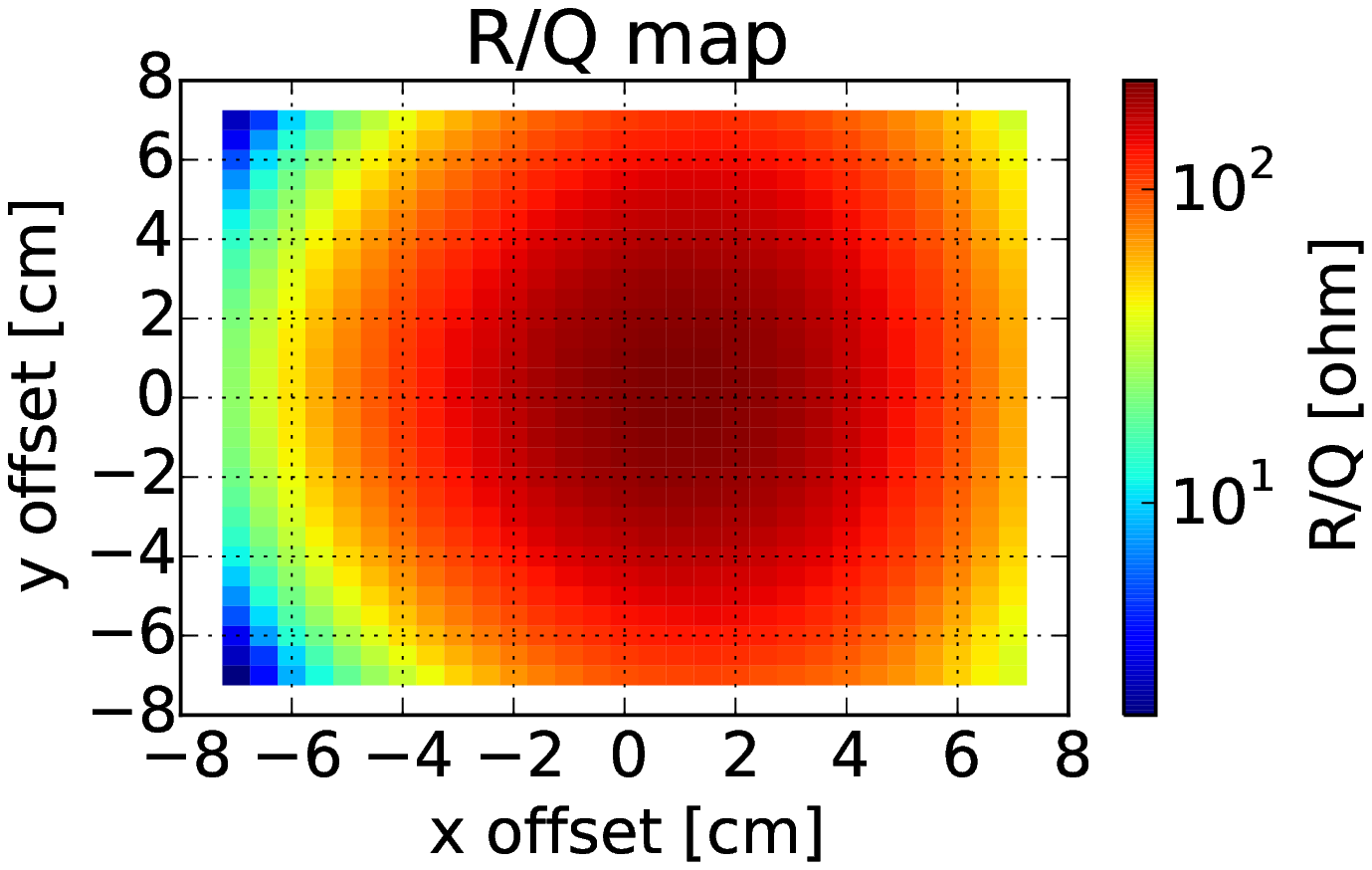}}  
\hspace{0em}
  \subcaptionbox{\label{fig:ellip_mode1e}M1: ellip. TM$_{010}$}{\includegraphics[width=0.3\textwidth]{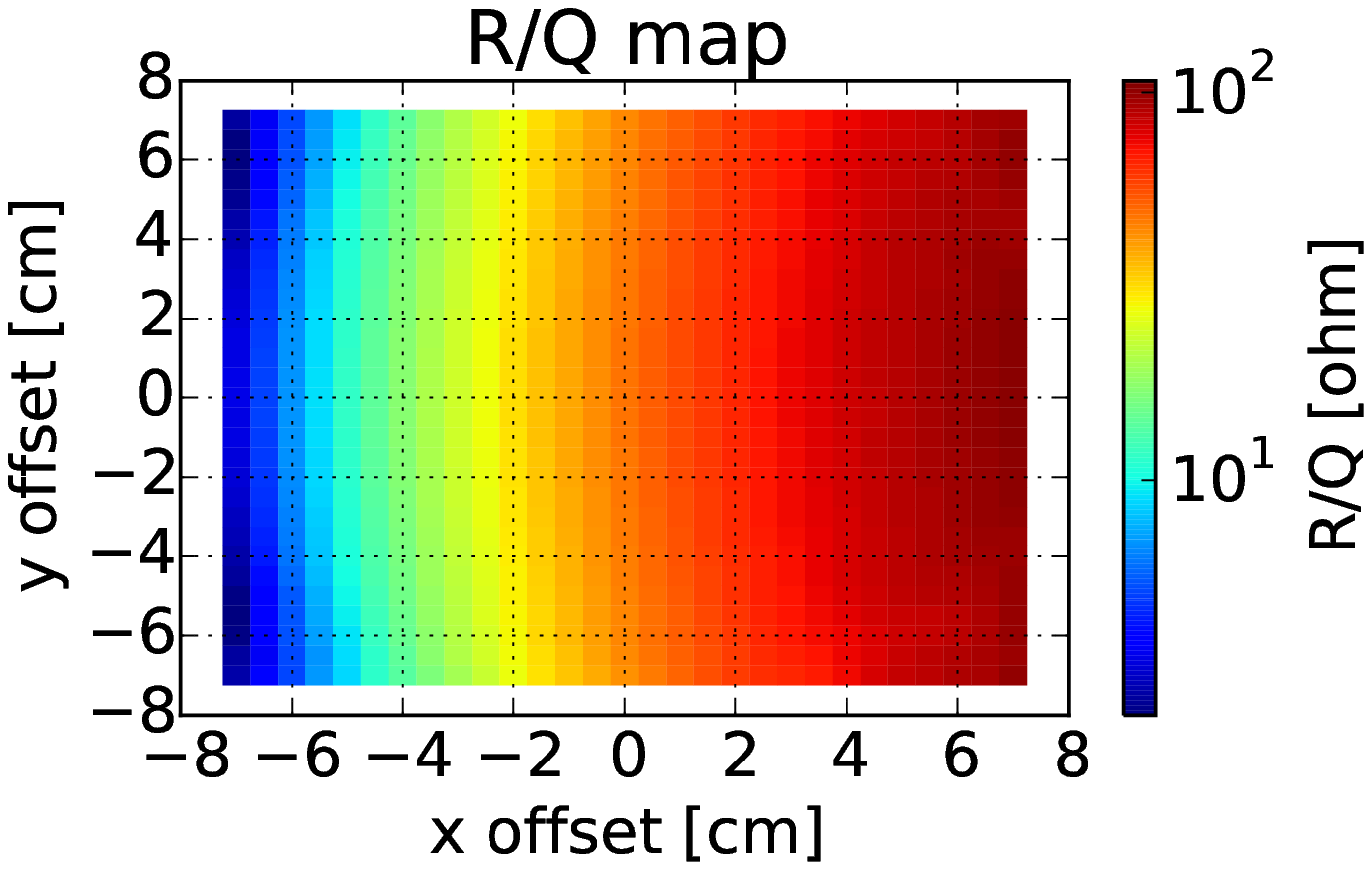}}  
  \subcaptionbox{\label{fig:ellip_mode2e}M2: ellip. TM$_{110}$}{\includegraphics[width=0.3\textwidth]{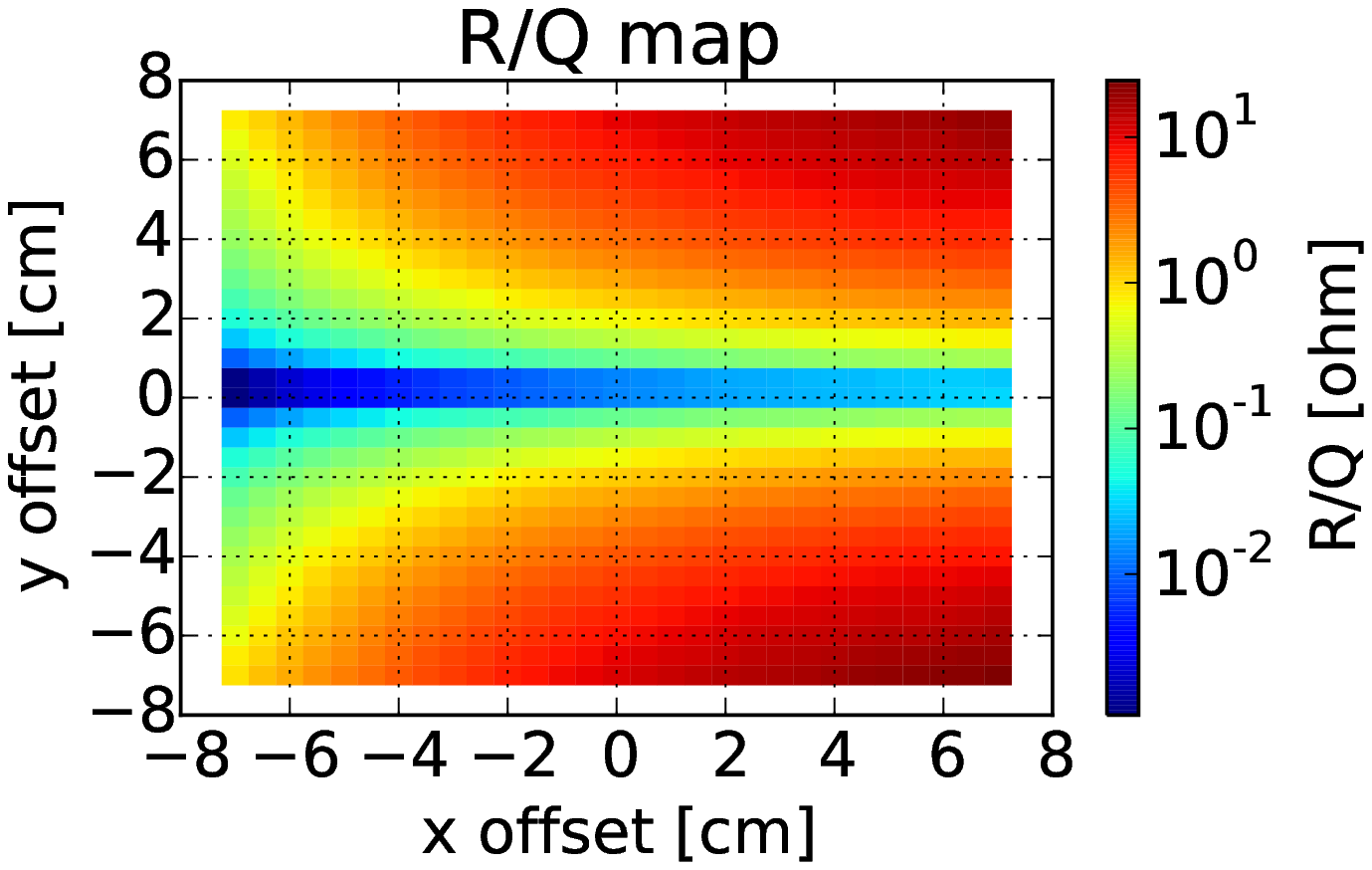}}  
    \subcaptionbox{\label{fig:ellip_mode3e}M3: ellip. TM$_{210}$}{\includegraphics[width=0.3\textwidth]{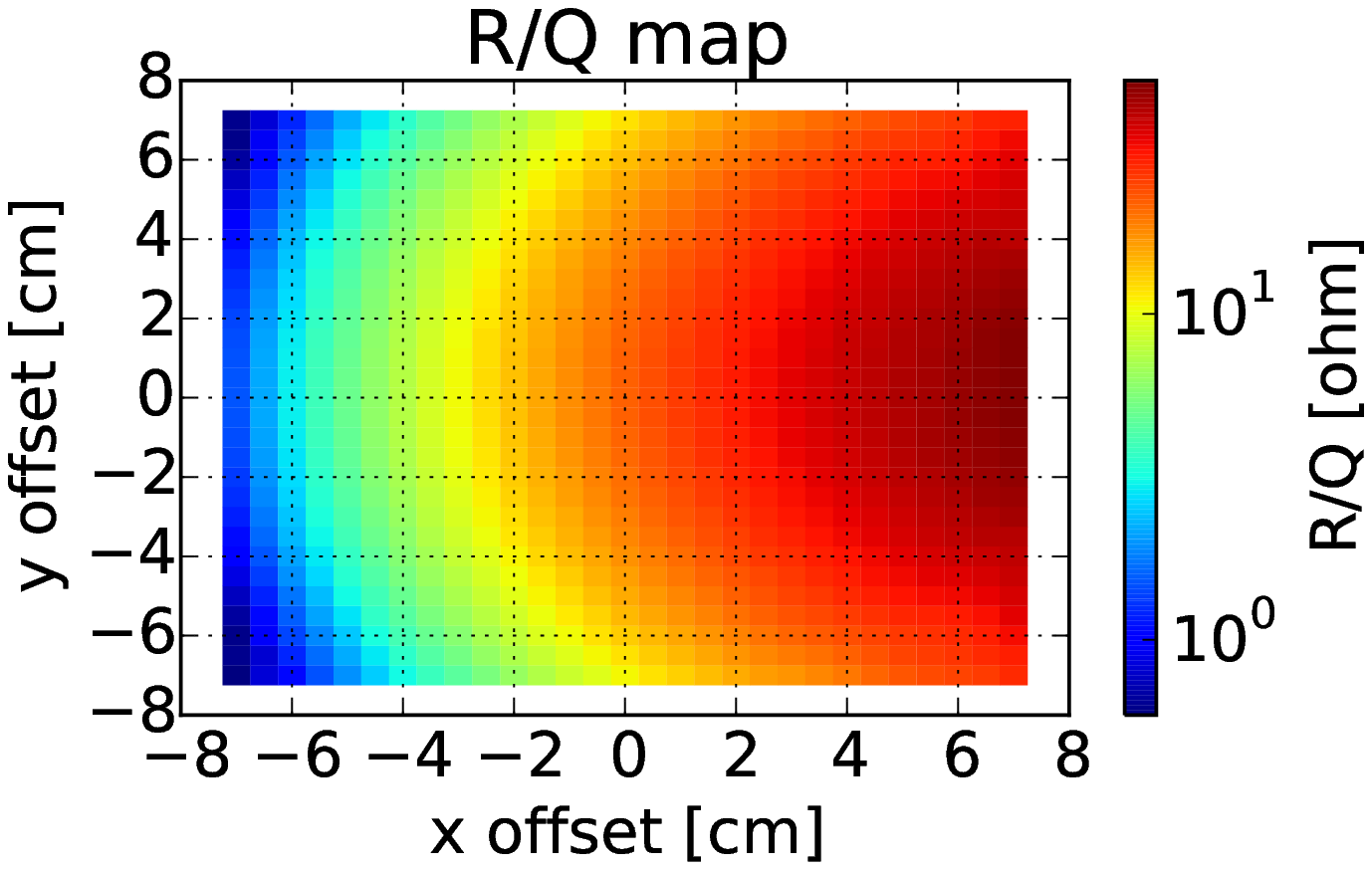}}  
\caption{R/Q versus position for the first \colour{three modes M1, M2 and M3} from left to right, around the center of the beam pipe. Top row circular, bottom row elliptical. Color scale for R/Q is logarithmic. (color)}
\label{fig:fig}
\end{figure}

One of the geometries that allows for the above mentioned R/Q map is an elliptical shallow pillbox working in its TM010 mode. The bound electromagnetic fields differ from those of circularly cylindrical resonators \cite{yeh1962}. By contrast to circular shapes the fall-off of the amplitude of the $E_z$ component of the standing wave along the beam axis is more dramatic along the minor axis of the ellipse, compared to the one along the major axis. By putting the beam pipe off center one can make use of the one-sided steep fall-off while the two-sided slow fall-off along the large axis is even favourably more flattened by the presence of the large aperture beam pipe. In figure \ref{fig:fig} the R/Q map is shown as a colour map showing the mode impedance versus transversal position offset using Microwave Studio\textsuperscript{\textregistered}. The results in the top row of this figure are obtained by simulating a circular pillbox cavity of radius 20 cm. The center of the beam pipe is offset to $r/2=10$ cm. The beam pipe has a radius of 9.5 cm. The bottom row shows the same configuration, but a for an elliptical cavity whose major axis is 3 times the minor axis, whereas the beam pipe is still offset halfway along the minor axis (see figure \ref{fig:dimensions}). The corresponding frequencies are listed in table \ref{tab:tab}.
\begin{table}
\centering
\caption{\label{tab:tab}Eigenmode frequency and unloaded Q values$^{\dagger}$ \colour{of the first four modes.}}
\footnotesize
\begin{tabular}{@{}lllll}
\br
Mode&\multicolumn{2}{c}{Circular} &\multicolumn{2}{c}{Elliptical}\\
&Freq [MHz]&Q&Freq [MHz]&Q\\
\mr
M1& 582.787 & 27630 & 422.838 & 23381\\
M2& 901.159&33698 & 509.778 & 25580 \\
M3& 954.758&37462 & 614.954 & 28984\\
M4 & 1142.481 & 64126 & 708.236 & 31812\\ 
\br
\end{tabular}\\
$^{\dagger}$Calculated in perturbation mode using a perfect copper surface with $\sigma = 5.8\times 10^7$ S/m.
\end{table}
\normalsize

\begin{figure}[htb]
\centering
  \centering
  \subcaptionbox{\label{fig:ellip_mode1e_roqx}}{\includegraphics[width=0.49\textwidth]{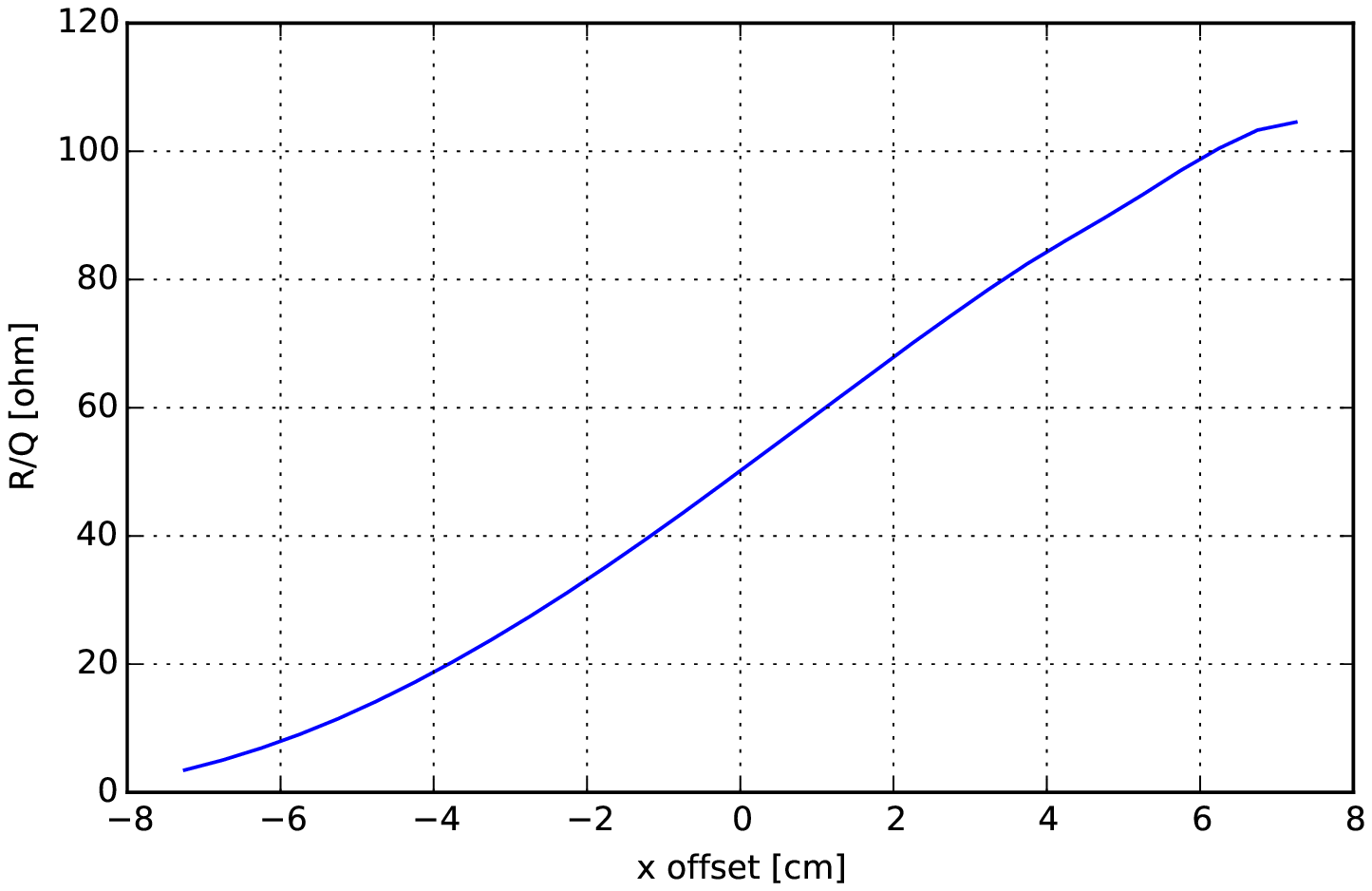}}  
  \subcaptionbox{\label{fig:ellip_allmode_efieldy}}{\includegraphics[width=0.49					\textwidth]{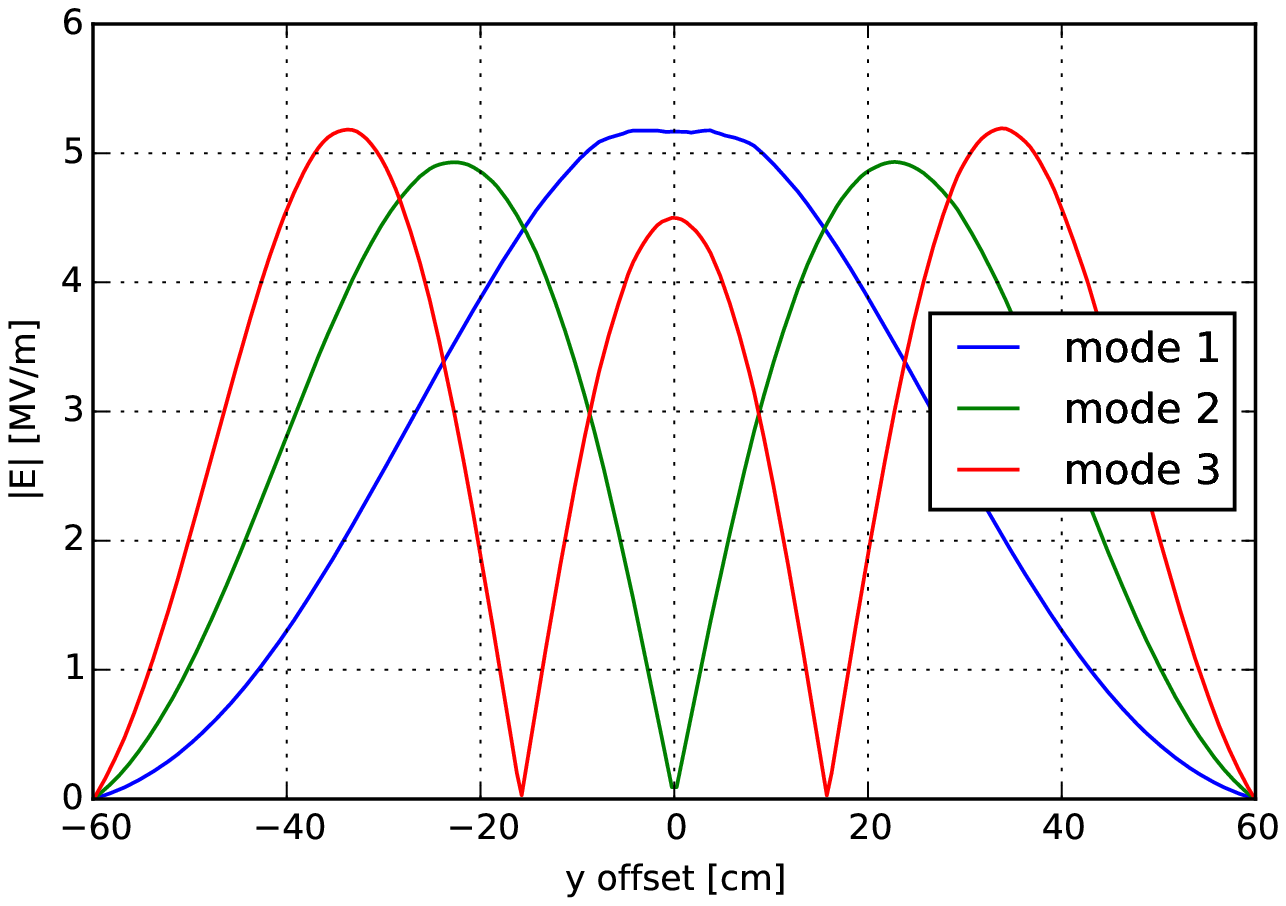}}  
\caption{Simulated longitudinal R/Q along the minor axis (left) and field amplitude along the major axis (right) whereas the field energy is normalized to 1 J (color).}
\label{fig:fig2}
\end{figure}

This approach has some potential advantages over existing approaches to design circular cavity based beam position monitors that utilize a circular dipole mode as shown in figure \ref{fig:circ_mode2e}. In such designs, the dipole mode needs to be extracted and often the presence of a strong fundamental monopole (see e.g. figure  \ref{fig:circ_mode1e}) mode is a problem. In the present design the strong fundamental monopole mode itself is used (figure \ref{fig:ellip_mode1e}) which results in \colour{a high sensitivity}. The fundamental mode of the elliptical design stretches out in the direction of major axis, allowing for a large area where the R/Q is decreasing almost linearly along the x direction (see figure \ref{fig:ellip_mode1e_roqx}). Simultaneously this area of interest is nearly cleared from the unwanted fields of the dipole mode. For instance, the value of R/Q at y=-2 cm in the elliptical dipole mode as shown in figure \ref{fig:ellip_mode2e} shows a reduction of approximately one order of magnitude compared to that of the circular mode shown in figure \ref{fig:circ_mode2e}.

The flatness of the R/Q map should be maximized for a given set of parameters imposed by the storage ring, such as mechanical restrictions and beam pipe aperture at the position of installation, which is preferably where the dispersion is maximum.

\subsection{Coupling mechanisms}

The stretch in one transversal direction comes at a cost of close lying higher order modes (HOM) (see table \ref{tab:tab}). This is of little concern in cavity pick-ups designed to have high sensitivities to detect a few charged particles. Also, frequencies above the beam pipe cut-off frequency (here 925 MHz for TE11) will not be trapped. Still, due to the asymmetric design, any remaining HOM fields can be selectively damped using HOM couplers. The elliptical design fortunately allows for a spatial separation (see figure \ref{fig:ellip_allmode_efieldy}), where electrical antennas can be placed on the face along the major axis pointing into z direction. The places of interest are the maxima of the e-field of the modes in figure \ref{fig:ellip_allmode_efieldy}. For the fundamental mode, while the signal can also be coupled electrically, a magnetic loop is much more desirable to better control the coupling constant which in turn directly affects the loaded Q value.

\subsection{Mode of operation}

The information on position is hidden in the amplitude of the coupled signal. So by contrast to traditional beam position monitors, this design can determine position of bunched as well as coasting beams. After normalizing the signal to that of a circular reference cavity, this signal needs to be mixed down and passed through a level-detector for further use. Additionally, due to the position dependence of the R/Q, standard techniques can be used to extract information on tune and chromaticity \cite{caspers2009}.

\begin{figure}[htb]
\centering
  \centering
  \subcaptionbox{\label{fig:dimensions}}{\includegraphics[width=0.15\textwidth]{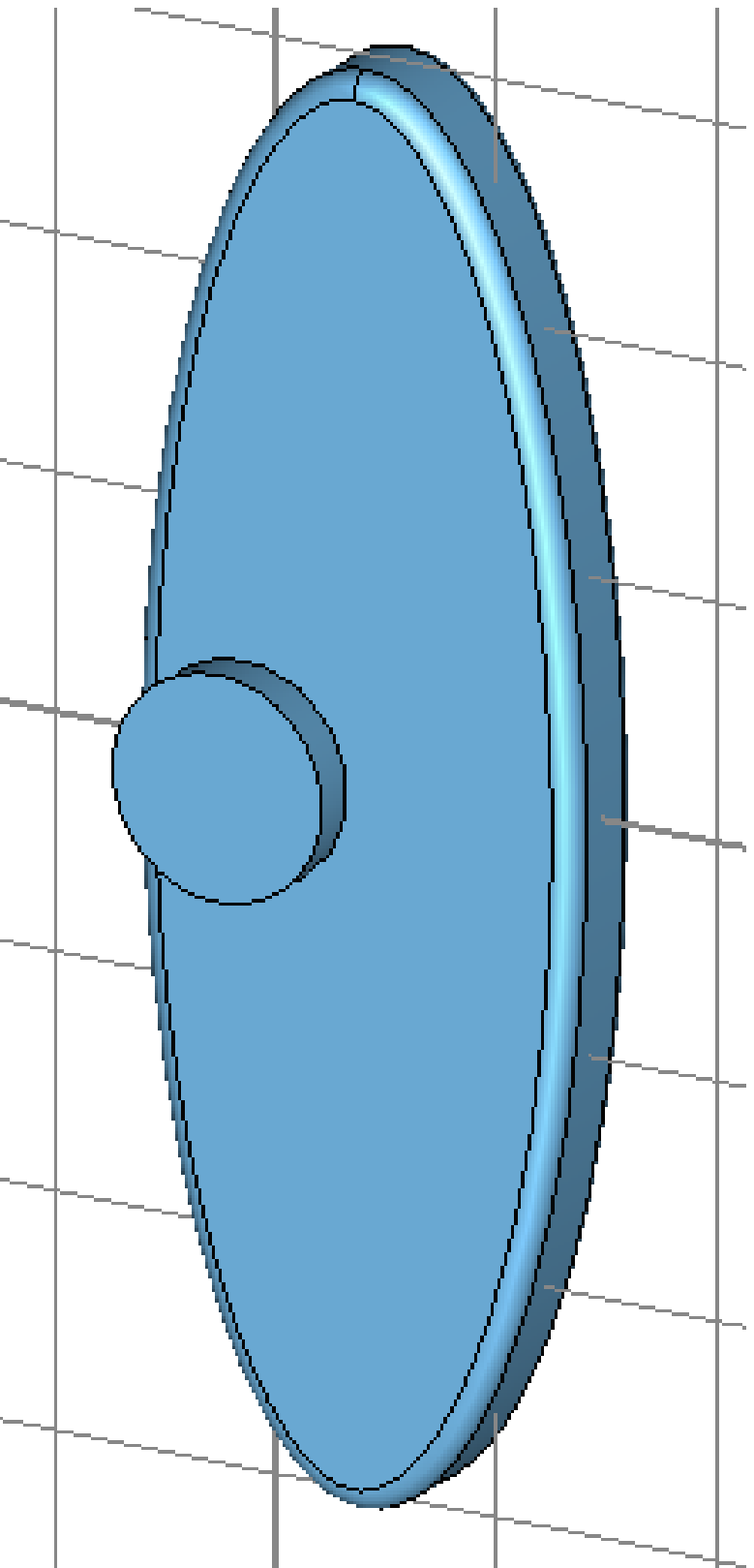}} 
  \hspace{2cm}
    \subcaptionbox{\label{fig:ellip_cast}}{\includegraphics[width=0.25\textwidth]{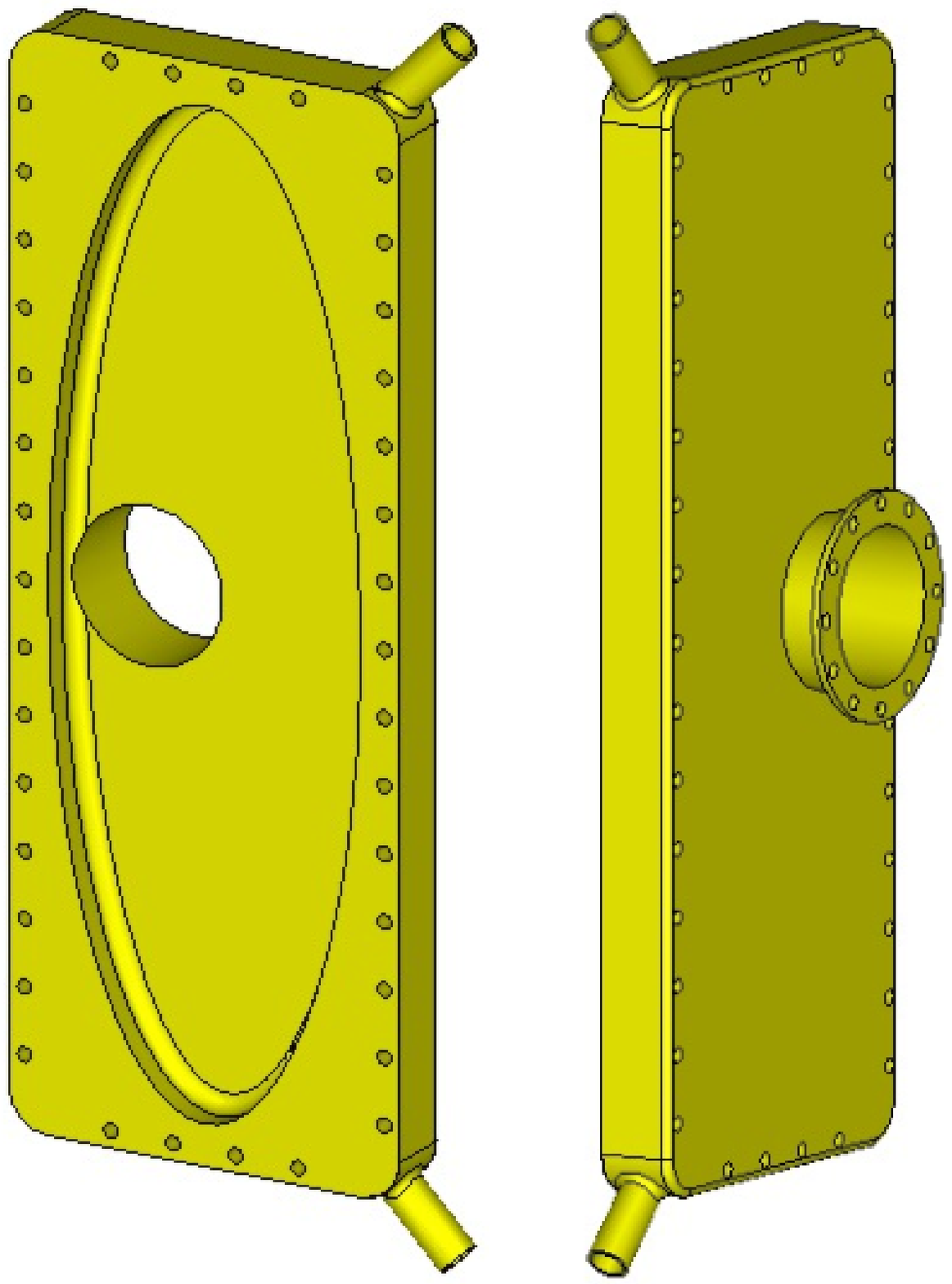}}  
\caption{The simulated cavity with an elliptical cross section (left) and possible CAD Model (right). Minor and major axis, depth and pipe radius are 20, 60, 10 and 9.5 cm respectively (color).}
\label{fig:fig2}
\end{figure}

Finally, in order to tune the frequency of the cavity to exactly match a desired Schottky harmonic, remotely controlled plungers are needed, which can be installed symmetrically as shown in figure \ref{fig:ellip_cast}.

\section{Conclusion and outlook}

This work presents a novel approach to designing a highly sensitive resonant cavity pickup for non-destructive detection of intensity and position of ion beams in storage rings. By using an elliptical geometry, limitations rising from large beam pipe aperture can be circumvented. Future challenges are optimizations with regard to specific beam parameters in the CR storage ring of the FAIR project such as the isochronous ion-optical mode and the position information of cocktail beams \cite{walker2013}. A bench-top model of a circular pillbox cavity with offset beam pipes has been constructed as a toy model for the verification of the R/Q map. To this end, a fully automated test-bench has been constructed in order to measure this and future model cavities using bead pull perturbation method. A complete account can be found in \cite{chen2014}.

\ack
M.S.S. acknowledges partial support by HA216/EMMI. X.C. acknowledges support of the EU contract No. PITN-GA-2011-289485. This work is partially supported by Helmholtz-CAS Joint Research Group HCJRG-108. The authors thank I. Schurig, F. Caspers, P. Kowina, S. A. Litvinov and A. Mostacci for fruitful discussions.

\section*{References}


\end{document}